\documentclass[12pt,thmsa]{article}
\usepackage{amsfonts}

\input{tcilatex}
\begin{document}

\author{Karl-Georg Schlesinger \\
Institute for Theoretical Physics\\
Vienna University of Technology\\
Wiedner Hauptstrasse 8-10/136\\
1040 Vienna, Austria\\
e-mail: kgschles@esi.ac.at}
\title{Some remarks on mirror symmetry and noncommutative elliptic curves}
\date{}
\maketitle

\begin{abstract}
This paper agrees basically with the talk of the author at the workshop
``Homological Mirror Symmetry and Applications'', Institute for Advanced
Study, Princeton, March 2007.
\end{abstract}

\section{Introduction}

This paper starts from the question of how to extend the well known mirror
symmetry of elliptic curves (see \cite{Dij}) - which was a motivating guide
for the formulation of the homological mirror symmetry conjecture in \cite
{Kon} - to the noncommutative case. We do not claim to perform this
extension, here, but restrict to a few simple remarks on the subject.

We start in the next section with two examples which motivate the
consideration of noncommutative extensions of mirror symmetry. The second
example even shows a case where such an extension is not an option but
necessary. Besides this, the second example can - in a simple special case -
directly be reduced to the question of a noncommutative extension of mirror
symmetry for elliptic curves. In section 3, we collect some of the needed
results on mirror symmetry of (commutative) elliptic curves and in section 4
we consider the case of noncommutative elliptic curves.

We will suggest a definition of Gromov-Witten invariants for noncommutative
elliptic curves by a suitable generalization of a fermion partition function
on elliptic curves. Besides this, we will argue that the bosonic version of
this theory - which in the classical case corresponds to the Kodaira-Spencer
theory formulation on the mirror - might correspond to a 12-dimensional
theory with a cubic interaction term.

\bigskip

\section{Motivation}

Let us briefly discuss two examples which serve as motivation for the
question of how mirror symmetry extends to the case of noncommutative
manifolds.

The first example comes from the work of \cite{KW}. There, the geometric
structures and the statement of the geometric Langlands program for
algebraic curves are derived from the $S$-duality conjecture for $N=4$ SUSY
YM-theory in $d=4$ by reducing a topologically twisted form of the four
dimensional gauge theory (with gauge group $G$) to a two dimensional sigma
model with the target space given by the Hitchin moduli space $Hit\left(
G,C\right) $ of the compact Riemann surface $C$ - the compactification space
in the dimensional reduction - and the gauge group $G$. The topologically
twisted gauge theory contains a - so called canonical - parameter $\Psi $
which combines the coupling parameter $\tau $ of the SUSY\ YM-theory 
\[
\tau =\frac \theta {2\pi }+\frac{4\pi i}{e^2} 
\]
(where $\theta $ is the $\theta $-angle and $e$ gives the gauge coupling)
with the twisting parameter $t$ (which parametrizes the family of possible
topological twistings of the type used in \cite{KW}) as 
\[
\Psi =\frac{\tau +\overline{\tau }}2+\frac{\tau -\overline{\tau }}2\left( 
\frac{t-t^{-1}}{t+t^{-1}}\right) 
\]
The two sides of the geometric Langlands duality arise at parameter values $%
\Psi =0$ and $\Psi =\infty $, respectively.

One can now pose the question of what happens at a more general parameter
value $\Psi \in \Bbb{C}$. For $\Psi \in \Bbb{R}$ one may always assume $t=1$
and hence 
\[
\Psi =\frac \theta {2\pi }=\func{Re}\left( \tau \right) 
\]
With $\omega _I$ denoting the symplectic form corresponding to the $I$
choice of complex structure on the Hyperk\"{a}hler manifold $Hit\left(
G,C\right) $ (see \cite{KW} for the details), the 2d sigma model in this
case contains a $B$-field 
\[
B=-\omega _I\func{Re}\left( \tau \right) 
\]
So, for more general values of $\Psi $ a $B$-field arises and one gets a
twisted version of geometric Langlands duality (quantum geometric Langlands
duality, see section 11.3 of \cite{KW}). Since the $S$-duality of the 4d
gauge theory reduces to (homological)\ mirror symmetry for the 2d sigma
model, the question for a noncommutative extension of mirror symmetry
arises, here.

\bigskip

As a second example, let us consider the case of the $D5$-brane worldvolume
gauge theory in type IIB string theory on a - possibly singular -- $K3$%
-surface $X$ (or $X=T^4$). This theory appears as the effective limit of
little string theory (LST)\ in type IIB. One can show that the gauge group
of the LST - and of the 6d effective field theory - has to belong to one of
the three $ADE$-series. We will completely restrict to the $U\left( 1\right) 
$-case in the $A$-series and to $X=T^4$, here.

Concretely, let $M$ be the worldvolume of a $D5$-brane in type IIB string
theory, $F$ the curvature of a $U\left( 1\right) $-connection of a line
bundle over $M$, $B$ the $NS$ 2-form field. Let $\mathcal{C}$ be the
background $RR$ gauge field, i. e. 
\[
\mathcal{C}=\theta +\widetilde{B}+G 
\]
with $\theta $ a scalar, $\widetilde{B}$ the $RR$ 2-form field and $G$ a
4-form field with self-dual field strength 
\[
dG=*dG 
\]
Let $v$ be the $RR$ charge vector (Mukai vector) 
\[
v=Tr\ \exp \left( \frac{iF}{2\pi }+B+\frac{c_2}{24}\right) 
\]
With 
\[
\mathcal{F}=F-2\pi iB 
\]
the action of the effective 6d CFT can be written as (see \cite{Dij 1998}
for more details) 
\begin{equation}
S=\int_M\frac 1{g_s}Tr\ \mathcal{F}\wedge *\mathcal{F}+\mathcal{C}\wedge
v\left( \mathcal{F}\right)  \label{6}
\end{equation}
Consider now the case 
\[
M=\Sigma \times X 
\]
with $\Sigma $ a Riemann surface (or non-compact) and $X=T^4$ with 
\[
vol\left( X\right) \ll vol\left( \Sigma \right) 
\]
In the limit of small $X$, (\ref{6}) can, again, be dimensionally reduced to
a sigma model on $\Sigma $ with the target space given by $Inst\left(
X\right) $, the instanton moduli space (anti-self-dual connections $F_{+}=0$%
) on $X$. For $X$ Hyperk\"{a}hler (i.e. $X=T^4$ or $X$ a K3 surface), $%
Inst\left( X\right) $ is a - singular - Hyperk\"{a}hler manifold.

In the $U\left( 1\right) $-case, the singular Hyperk\"{a}hler structure of $%
Inst\left( X\right) $ can be regularized to a smooth Hyperk\"{a}hler
manifold by the large $N$ limit of the Hilbert scheme of $N$ points on $X$.
On the other hand, the Hilbert scheme of $N$ points on $X$ also regularizes
the orbifold $S^NX$, the $N$-fold power of $X$ modulo the action of the
symmetric group.

Let us pose the question if there exists a canonical coisotropic brane (c.c.
brane), i.e. a target space filling coisotropic brane, for this
Hyperk\"{a}hler sigma model. In the case of the Hitchin moduli space sigma
model, the c.c. brane is used in \cite{KW} to derive the $D$-module property
of the Hecke eigensheaves, i.e. it is essential to derive the structures of
the geometric Langlands program from $S$-duality (also the existence of the
c.c. brane is believed - \cite{Wit} - to be essential to derive the 2d
conformal field theory approach to the geometric Langlands program - see 
\cite{Fre} - from the setting of \cite{KW}).

In \cite{Dij 1998} it was shown that the $RR$-fields of (\ref{6}) contribute
to the $NS$ 2-form field of the $Inst\left( X\right) $ sigma model under
dimensional reduction while both the K\"{a}hler form and the $NS$ 2-form $B$%
-field on $X$ in (\ref{6}) are used to determine the K\"{a}hler form on $%
Inst\left( X\right) $, i.e. the Hyperk\"{a}hler structure on $Inst\left(
X\right) $ is not well defined unless one specifies the $B$-field on $X$.

The condition for the existence of a coisotropic $A$-brane of full dimension
(c.c. brane) in a sigma model with target $Y$ was shown in \cite{KO} to be: 
\begin{equation}
\left( \omega ^{-1}\widetilde{F}\right) ^2=-1  \label{7}
\end{equation}
with $\widetilde{F}$ the curvature of the connection of the bundle defining
the c.c. brane and $\omega $ the symplectic form on $Y$.

Assume, now, that the $NS$ 2-form field $B$ in (\ref{6}) vanishes. In this
case (see \cite{Dij 1998}), the K\"{a}hler structure of $Inst\left( X\right) 
$ is given by the large $N$ limit of $S^NX$, i.e. the K\"{a}hler structure
is determined by the symmetric $N$-fold products of the K\"{a}hler structure
of $X$. But for $X=T^4$ and generic $\omega $, it was argued in \cite{KO}
that a coisotropic brane of full dimension should be impossible. This
argument can be adapted to $S^NX$. This means that generically there should
exist no c.c. brane on $Inst\left( X\right) $ and the sigma model defined by
the dimensional reduction of (\ref{6}) should therefore radically differ in
this respect from the setting of \cite{KW}.

Up to now, we have implicitly assumed that the $RR$ background gauge fields
of (\ref{6}) vanish, i.e. the $Inst\left( X\right) $ sigma model has a
vanishing $NS$ 2-form field $\widehat{B}$. Let us now assume that $\widehat{B%
}\neq 0$. But we have 
\[
d\widehat{B}=0 
\]
We consider the large volume limit of $Inst\left( X\right) $ (corresponding
to small string coupling $g_s$, see \cite{Dij 1998}).

Let $\widetilde{\mathcal{F}}$ be the curvature $\widetilde{F}$ shifted by $%
\widehat{B}$, i.e. 
\[
\widetilde{\mathcal{F}}=\widetilde{F}-2\pi i\widehat{B}
\]
Equation (\ref{7}) should then be replaced by 
\begin{equation}
\left( \omega ^{-1}\widetilde{\mathcal{F}}\right) ^2=-1  \label{8}
\end{equation}
It can be shown that the field $\widehat{B}$ can always be fine-tuned such
that a c.c. brane exists (i.e. the $RR$ background field can always be
fine-tuned to allow for the existence of a c.c. brane). Here, one makes use
of the fact that the components of the $NS$ 2-form field of the $Inst\left(
X\right) $ sigma model induced from the $RR$ background gauge fields of (\ref
{6}) constitute a basis of $H^2\left( Inst\left( X\right) ,\Bbb{R}\right) $
(see \cite{Dij 1998}). 

Actually, the components of $\widehat{B}$ induced from $\theta $ and $%
\widetilde{B}$ already constitute a basis of $H^2\left( Inst\left( X\right) ,%
\Bbb{R}\right) $ while we have the following additional condition on the $RR$
background gauge fields (see \cite{Dij 1998}): If 
\[
v\in \left( Q_5,Q_3,-Q_1\right) \in H^{*}\left( X,\Bbb{Z}\right) 
\]
is the Mukai vector, we have 
\[
v\cdot \mathcal{C}=Q_1\cdot \theta +Q_3\cdot \widetilde{B}-Q_5\cdot G=0 
\]
So, if $G$ would vanish, we would have an additional relation between $%
\theta $ and $\widetilde{B}$, violating the basis property of the $\widehat{B%
}$-field contributions induced by them. In consequence, $G$ can not vanish.

We can therefore draw the following conclusion: For non-vanishing $RR$
background 4-form field $G$ in (\ref{6}) - and therefore for non-vanishing $%
NS$ 2-form field $\widehat{B}$ in the $Inst\left( X\right) $ sigma model - 
there exists always a c.c. brane in the $Inst\left( X\right) $ sigma model.

In consequence, if we want to study mirror symmetry in the sigma model
reduction of (\ref{6}) in the presence of a c.c. brane, we once again arrive
at the question of a noncommutative extension of mirror symmetry. Observe
that in this case the noncommutative extension is not just an option for a
generalization but is necessary since for $\widehat{B}=0$ a c.c. brane does
not exist generically.

As a special case, one can show that one can choose a $\widehat{B}$ which is
induced from a 2-form field on $X=T^4$ on the symmetric powers $S^NX$. We
can make an even more special choice by requiring that the 2-form field on $%
T^4$ should respect the factorization 
\[
T^4\cong T^2\times T^2
\]
and be constant. In consequence, we can study the effect of non-vanishing
4-form field $G$ in (\ref{6}) in a special case by studying a field $%
\widehat{B}$ which is induced from a simple constant 2-form field on an
elliptic curve.

In other words, in the simple case of such a factorizable field $\widehat{B}$%
, we can study the question of a noncommutative extension of mirror symmetry
in the sigma model reduction of (\ref{6}) by starting from the question of a
noncommutative extension of mirror symmetry for elliptic curves. It is this
question which forms the topic of the present paper. We do not claim to
present a noncommutative extension of mirror symmetry for elliptic curves,
here, but restrict to a few small remarks on the subject.

\bigskip

\section{The elliptic curve}

Let us start by very briefly reviewing the case of mirror symmetry for
(commutative)\ elliptic curves (see \cite{Dij} and references cited therein,
especially \cite{Dou}, \cite{KaZa}, \cite{Rud}).

An elliptic curve $E_{t,\tau }$ is a smooth 2-torus equipped with a
holomorphic and a symplectic structure. The holomorphic structure -
parametrized by $\tau $ - is given by the representation of the elliptic
curve as 
\[
\Bbb{C}/\left( \Bbb{Z}\oplus \Bbb{Z}\tau \right) 
\]
with $\tau =\tau _1+i\tau _2\in \Bbb{C}$ from the upper half plane $\Bbb{H}$%
, i..e $\tau _2>0$. The symplectic structure - parametrized by $t\in \Bbb{H}$
- is given by the complexified K\"{a}hler class $\left[ \omega \right] \in
H^2\left( E_{t,\tau },\Bbb{C}\right) $ with 
\[
\omega =-\frac{\pi t}{\tau _2}dz\wedge d\overline{z} 
\]
and for $t=t_1+it_2$ the area of the elliptic curve is given by $t_2$.
Mirror symmetry relates the elliptic curves $E_{t,\tau }$ and $E_{\tau ,t}$.

On the symplectic side, we have the Gromov-Witten invariants $F_g$, defined
as the generating functions for counting $d$-fold connected covers of $%
E_{t,\tau }$ in genus $g$. One can combine the functions $F_g$ into a
two-variable partition function 
\[
Z\left( q,\lambda \right) =\exp \sum_{g=1}^\infty \lambda ^{2g-2}F_g\left(
q\right) 
\]
with $q=e^{2\pi it}$.

Now, it is important that $Z\left( q,\lambda \right) $ can be calculated in
three different ways (see Theorem 1 - Theorem 3 of \cite{Dij}). The first
case is a large $N$ calculation in terms of $U\left( N\right) $ Yang-Mills
theory on $E_{t,\tau }$. We will not refer to this case, here. The second
possibility is a calculation in terms of a Dirac fermion on the elliptic
curve. Starting from Dirac spinors $b,c$ on the elliptic curve with action 
\[
S=\int_{E_{t,\tau }}\left( b\overline{\partial }c+\lambda b\partial
^2c\right) 
\]
one shows that the operator product expansion defines a fermionic
representation of the $W_{1+\infty }$ algebra. The partition function can be
calculated as a generalized trace (as defined in \cite{AFMO}) of this
algebra, leading to 
\begin{equation}
Z\left( q,\lambda \right) =q^{-\frac 1{24}}\oint \frac{dz}{2\pi iz}%
\prod_{p\in \Bbb{Z}_{\geq o}+\frac 12}\left( 1+zq^pe^{\lambda p^2}\right)
\left( 1+\frac 1zq^pe^{-\lambda p^2}\right)  \label{1}
\end{equation}
(see \cite{Dij 1996} for the details). Note that for the action and the
partition function above - and for the sequel of this paper - we have
changed the notation to denote the parameter values of the mirror elliptic
curve by $t$ and $\tau $. It is this representation of $Z\left( q,\lambda
\right) $ which leads to the famous theorem of Dijkgraaf, Kaneko, Zagier
stating that the functions $F_g\left( q\right) $ are quasi-modular forms
(i.e. $F_g\in \Bbb{Q}\left[ E_2,E_4,E_6\right] $ where $E_2,E_4,E_6$ are the
classical Eisenstein series of weight 2, 4, and 6, respectively) and have
weight $6g-6$.

Finally, as in the case of Calabi-Yau 3-folds, by mirror symmetry $Z\left(
q,\lambda \right) $ can be calculated as the partition function of a
Kodaira-Spencer theory. In the case of elliptic curves, this is given by the
action of a simple real bosonic field with $\left( \partial \varphi \right)
^3$ interaction term, i.e. by the action 
\[
S\left( \varphi \right) =\int_{E_{\tau ,t}}\left( \frac 12\partial \varphi 
\overline{\partial }\varphi +\frac \lambda 6\left( -i\partial \varphi
\right) ^3\right) 
\]
(see \cite{Dij}, \cite{Dij 1996} for the details).

\bigskip

\section{The noncommutative elliptic curve}

Let us now come to the question how mirror symmetry and the above results
might generalize to the noncommutative torus. We will start with the case of
the fermionic representation of $Z\left( q,\lambda \right) $. The first
question we have to face is how the full structure of an elliptic curve,
beyond the structure of a smooth torus, generalizes to the noncommutative
case. Holomorphic structures on the noncommutative torus have been
introduced in \cite{Pol 2003}, \cite{Pol 2004}, \cite{Pol 2005}, and \cite
{PS}. Unfortunately, $Z\left( q,\lambda \right) $ is not expressed in terms
of a single elliptic curve but in terms of the modular parameter $q$ of the
whole family of elliptic curves. So, to arrive at an analogue of (\ref{1}),
we have to consider an extension of the range of the modular parameter,
including noncommutative elliptic curves. In \cite{Soi} it is shown that one
can view the noncommutative torus as the degenerate limit $\left| q\right|
\rightarrow 1$ of classical elliptic curves (observe that since $t\in \Bbb{H}
$ and $q=e^{2\pi it}$, $\left| q\right| <1$ for classical elliptic curves),
arriving in this way also at the notion of a noncommutative elliptic curve.
We will therefore discuss the question of a noncommutative analogue of (\ref
{1}) in the form of the question of performing the limit $\left| q\right|
\rightarrow 1$ in (\ref{1}).

Obviously, we can not directly perform the limit in (\ref{1}). Besides this,
there exist only very few results on $q$-analysis for $\left| q\right| =1$.
But fortunately there exists an elliptic deformation of the $q$-deformed
gamma function and this elliptic gamma function (which has two deformation
parameters)\ allows to take a limit in which a single unimodular deformation
parameter arises (\cite{Rui 1997}, \cite{Rui}). In this sense, the elliptic
gamma function includes the $q$-gamma function case with $\left| q\right| =1$%
. We will therefore consider the problem of taking the limit $\left|
q\right| \rightarrow 1$ in (\ref{1}) in the more general form of looking for
an elliptic analogue of (\ref{1}). We will proceed as follows: We will first
rewrite (\ref{1})\ in terms of $q$-deformed gamma functions (for the
classical case, i.e. $\left| q\right| <1$) and then replace these by the
elliptic gamma function of \cite{Rui 1997}, \cite{Rui}.

Let us start by considering the case $\lambda =0$. With the substitution $%
q\mapsto q^2$, we have 
\begin{eqnarray*}
Z\left( q,0\right) &=&q^{-\frac 1{12}}\oint \frac{dz}{2\pi iz}\prod_{j\geq
0}\left( 1+zq^{2j+1}\right) \left( 1+\frac 1zq^{2j+1}\right) \\
&=&q^{-\frac 1{12}}\oint \frac{dz}{2\pi iz}\left( -zq;q^2\right) _\infty
\left( -\frac qz;q^2\right) _\infty
\end{eqnarray*}
where 
\[
\left( a;q\right) _n=\prod_{j=0}^{n-1}\left( 1-aq^j\right) 
\]
is the $q$-shifted factorial and 
\[
\left( a;q\right) _\infty =\prod_{j=0}^\infty \left( 1-aq^j\right) 
\]
the limit $n\rightarrow \infty $ which exists for $\left| q\right| <1$.
Remember that the classical Jacobi theta function 
\[
\vartheta \left( z,q\right) =\sum_{n=-\infty }^{n=+\infty }z^nq^{n^2} 
\]
can be expressed in the form of the Jacobi triple product as 
\[
\vartheta \left( z,q\right) =\left( -zq;q^2\right) _\infty \left( -\frac
qz;q^2\right) _\infty \left( q^2;q^2\right) _\infty 
\]
i.e. $Z\left( q,0\right) $ is basically given by an integral over the first
two factors of $\vartheta \left( z,q\right) $.

Next, let us rewrite $\left( a;q\right) _\infty $ in terms of the function $%
\Gamma _q$ with 
\[
\Gamma _q\left( x\right) =\frac{q^{-\frac{x^2}{16}}}{\left( -q^{\frac
12\left( x+1\right) };q\right) _\infty } 
\]
Observe that this form of the $q$-deformed gamma function (as used e.g. in 
\cite{Sto}) differs slightly from the usually used $q$-gamma function 
\[
\gamma _q\left( x\right) =\frac{\left( q;q\right) _\infty }{\left(
q^x;q\right) _\infty }\left( 1-q\right) ^{1-x} 
\]
Solving 
\[
a=-q^{\frac 12\left( x+1\right) } 
\]
for $x$, we arrive at 
\[
x=\frac{\log \left( a^2\right) }{\log \left( q\right) }-1=2\frac{\log \left(
a\right) }{\log \left( q\right) }-1 
\]
and 
\[
\left( a;q\right) _\infty =\frac{q^{-\left( \frac{2\frac{\log \left(
a\right) }{\log \left( q\right) }-1}4\right) ^2}}{\Gamma _q\left( 2\frac{%
\log \left( a\right) }{\log \left( q\right) }-1\right) } 
\]
In consequence, we have 
\begin{equation}
Z\left( q,0\right) =q^{-\frac 1{12}}\oint \frac{dz}{2\pi iz}\frac{%
q^{^{-\left( \frac{\log \left( -z\right) }{2\log \left( q\right) }\right)
^2}}}{\Gamma _{q^2}\left( \frac{\log \left( -z\right) }{\log \left( q\right) 
}\right) \Gamma _{q^2}\left( -\frac{\log \left( -z\right) }{\log \left(
q\right) }\right) }  \label{2}
\end{equation}
Let for $q,p\in \Bbb{C}$ with $\left| q\right| ,\left| p\right| <1$ 
\begin{equation}
\Gamma \left( z;q,p\right) =\prod_{j,k=0}^\infty \frac{1-z^{-1}q^{j+1}p^{k+1}%
}{1-zq^jp^k}  \label{3}
\end{equation}
be the elliptic gamma function of \cite{Rui 1997}, \cite{Rui}. Then an
elliptic generalization of (\ref{2}) - which allows to take the limit to
unimodular $q$ in (\ref{2}) - would be 
\begin{eqnarray}
&&Z\left( q,p,0\right)  \label{4} \\
&=&q^{-\frac 1{12}}p^{-\frac 1{12}}\oint \frac{dz}{2\pi iz}\frac{%
q^{^{-\left( \frac{\log \left( -z\right) }{2\left( \log \left( q\right)
+\log \left( p\right) \right) }\right) ^2}}p^{^{-\left( \frac{\log \left(
-z\right) }{2\left( \log \left( q\right) +\log \left( p\right) \right) }%
\right) ^2}}}{\Gamma \left( \frac{\log \left( -z\right) }{\log \left(
q\right) +\log \left( p\right) };q^2,p^2\right) \Gamma \left( -\frac{\log
\left( -z\right) }{\log \left( q\right) +\log \left( p\right) }%
;q^2,p^2\right) }  \nonumber
\end{eqnarray}

Observe that $\Gamma \left( z;q,p\right) $ is symmetric in $q$ and $p$ which
guides our guess for the generalization of $Z\left( q,0\right) $.

Let us now discuss the case $\lambda \neq 0$. The factor 
\[
\left( -zq;q^2\right) _\infty =\prod_{j=0}^\infty \left( 1+zq^{2j+1}\right) 
\]
in $Z\left( q,0\right) $ is in this case deformed to 
\[
\prod_{j=0}^\infty \left( 1+zq^{2j+1}e^{\frac \lambda 2\left( j+\frac
12\right) ^2}\right) 
\]
Similarly, the factor 
\[
\left( -\frac qz;q^2\right) _\infty =\prod_{j=0}^\infty \left( 1+\frac
1zq^{2j+1}\right) 
\]
is deformed to 
\[
\prod_{j=0}^\infty \left( 1+\frac 1zq^{2j+1}e^{-\frac \lambda 2\left(
j+\frac 12\right) ^2}\right) 
\]
Since 
\[
\prod_{j=0}^\infty \frac{1-z^{-1}q^{j+1}}{1-zq^j}=\frac{\left(
z^{-1}q;q\right) _\infty }{\left( z;q\right) _\infty } 
\]
we make the following Ansatz for a generalization of the elliptic gamma
function (see (\ref{3})) to $\lambda \neq 0$: 
\[
\Gamma \left( z;q,p,\lambda \right) =\prod_{j,k=0}^\infty \frac{%
1-z^{-1}q^{j+1}p^{k+1}e^{-\frac \lambda 2\left( j+\frac 12\right)
^2}e^{-\frac \lambda 2\left( k+\frac 12\right) ^2}}{1-zq^jp^ke^{\frac
\lambda 2\left( j+\frac 12\right) ^2}e^{\frac \lambda 2\left( k+\frac
12\right) ^2}} 
\]
Let 
\[
\alpha _{q,p,\lambda }\left( z\right) =\frac{\log \left( -z\right) }{\log
\left( q\right) +\log \left( p\right) +\frac 94\lambda } 
\]
and 
\[
\widehat{\vartheta }\left( z;q,p,\lambda \right) =\frac{q^{-\frac{\alpha
_{q,p,\lambda }^2\left( z\right) }4}p^{-\frac{\alpha _{q,p,\lambda }^2\left(
z\right) }4}e^{-\frac 9{16}\lambda \alpha _{q,p,\lambda }^2\left( z\right) }%
}{_{\Gamma \left( \alpha _{q,p,\lambda }\left( z\right) ;q^2,p^2,\lambda
\right) \Gamma \left( -\alpha _{q,p,\lambda }\left( z\right)
;q^2,p^2,\lambda \right) }} 
\]

With these definitions at hand, we make the following Ansatz for a
generalization of $Z\left( q,p,0\right) $ to $\lambda \neq 0$: 
\begin{equation}
Z\left( q,p,\lambda \right) =q^{-\frac 1{12}}p^{-\frac 1{12}}\oint \frac{dz}{%
2\pi iz}\widehat{\vartheta }\left( z;q,p,\lambda \right)  \label{5}
\end{equation}
Of course, the numerical factors in the definition of $\widehat{\vartheta }%
\left( z;q,p,\lambda \right) $ are in no way unique. We have chosen a
definition, where the factors correspond to those appearing for $j=1$ in the
deformation of the $q$-shifted factorials appearing for $\lambda \neq 0$
(since this is how $q$ and $p$ appear in the $\lambda =0$ case in the
shifted factorials). We suggest (\ref{5}) as the definition for the
partition function of an elliptic fermion on the elliptic curve which
contains the degeneration to a single unimodular parameter (corresponding to
a fermion partition function on the noncommutative elliptic curve) as a
special case.

\bigskip

\begin{remark}
It is an open question for future research if (\ref{5}) corresponds for a
noncommutative elliptic curve to a fermionic action analogous to the action 
\[
S=\int_{E_{t,\tau }}\left( b\overline{\partial }c+\lambda b\partial
^2c\right) 
\]
of the commutative case.
\end{remark}

\bigskip

As in the classical case of commutative elliptic curves, we can use the
partition function (\ref{5}) to define Gromov-Witten invariants. Concretely,
in the classical case the definition of the partition function as 
\[
Z\left( q,\lambda \right) =\exp \left( \sum_{g=1}^\infty \lambda
^{2g-2}F_g\left( q\right) \right) 
\]
implies that we can calculate the Gromov-Witten invariants $F_g$ as 
\begin{equation}
F_g=\frac 1{\left( 2g-2\right) !}\frac{\partial ^{2g-2}\log \left( Z\right) 
}{\partial \lambda ^{2g-2}};_{\lambda =0}  \label{10}
\end{equation}
We can now use (\ref{10}), applied to the partition function (\ref{5}) as a
definition of elliptic Gromov-Witten invariants $F_g\left( q,p\right) $. The
limit to a single unimodular parameter can be used as a definition of
Gromov-Witten invariants for noncommutative elliptic curves.

\bigskip

Let us next consider the question of a noncommutative analogue of the
bosonic $\left( \partial \varphi \right) ^3$-action. We do not have
definitive results for this case but want to conclude this section with a
few remarks. The bosonic action has two properties which are decisive for
the calculation of Gromov-Witten invariants:

\bigskip

\begin{itemize}
\item  The interaction term is cubic.

\item  The interaction term is chiral.
\end{itemize}

\bigskip

Let us assume that mirror symmetry extends to the noncommutative case. More
concretely, let us assume that the partition function (\ref{5}) has a
representation by a bosonic action and that this action has (at least as one
contribution) a cubic chiral interaction term.

In the classical case, the bosonic representation is given by a real boson,
i.e. we have a real valued scalar field or more generally a section of a
line bundle. In the case of (\ref{5}), the integrand is mainly given by a
product of elliptic gamma functions. Now, it has been shown in \cite{FHRZ}, 
\cite{FV} that the elliptic gamma function is not related to a section of a
line bundle but to a section of a gerbe. One might therefore suspect that a
bosonic representation of (\ref{5}) - if it exists - should also be given in
terms of a bosonic field on a gerbe. So, one might suspect that locally the
bosonic field $\varphi $ is not given as a scalar but transforms as a 1-form
(remember that all fields transforming locally as a $p$-form are bosonic). $%
\partial \varphi $ should then be replaced by the differential on forms,
i.e. $\partial \varphi $ should locally transform as a 2-form. The
interaction term would then be (in order to be cubic) 
\[
\partial \varphi \wedge \partial \varphi \wedge \partial \varphi 
\]
and hence a 6-form. Since the interaction term should be chiral, we should
actually count degrees of forms in complex cohomology. Let us assume e.g.
that the interaction term transforms locally as a $\left( 0,6\right) $-form.
Analogous to the case of the $\left( 0,3\right) $-form in six dimensional
Kodaira-Spencer theory, we should integrate this together with a $\left(
6,0\right) $-form. In consequence, we arrive at the conclusion that the
bosonic theory - if it exists - should live on a 12-dimensional manifold.
So, we are lead to pose the following questions:

\bigskip

\begin{itemize}
\item  Does there exist a cubic twelve dimensional theory with the partition
function given by (\ref{5})?

\item  If yes, how is the twelve dimensional manifold determined?

\item  In six dimensional Kodaira-Spencer theory the field is an element of
a cohomology class, taking into account the gauge freedom of the field. One
would expect something similar to happen for the 2-form field $\partial
\varphi $. What is the correct type of cohomology theory?
\end{itemize}

\bigskip

We plan to come back to some of these questions in future work.

\bigskip

\textbf{Acknowledgments:} I would like to thank H. Grosse, A. Kapustin, M.
Kreuzer, and J. V. Stokman for discussions on or related to the material of
this paper.


\bigskip

\bigskip

\begin{thebibliography}{Dij 1994}
\bibitem[AFMO]{AFMO}  H. Awata, M. Fukuma, Y. Matsuo, S. Odake, \textit{%
Representation theory of the }W$_{1+\infty }$ \textit{algebra},
hep-th/9408158.

\bibitem[Dij 1994]{Dij}  R. Dijkgraaf, \textit{Mirror symmetry and elliptic
curves}, in \textit{The moduli space of curves}, Proceedings of the Texel
Island Meeting, April 1994, Birkh\"{a}user, Basel 1995.

\bibitem[Dij 1996]{Dij 1996}  R. Dijkgraaf, \textit{Chiral deformations of
conformal field theories}, hep-th/9609022.

\bibitem[Dij 1998]{Dij 1998}  R. Dijkgraaf, \textit{Instanton strings and
Hyperk\"{a}hler geometry}, hep-th/9810210.

\bibitem[Dou]{Dou}  M. R. Douglas, \textit{Conformal field theory techniques
in large N Yang-Mills theory}, hep-th/9311130.

\bibitem[FHRZ]{FHRZ}  G. Felder, A. Henriques, C. A. Rossi, C. Zhu, \textit{%
A gerbe for the elliptic gamma function}, math.QA/0601337.

\bibitem[Fre]{Fre}  E. Frenkel, \textit{Lectures on the Langlands program
and conformal field theory}, hep-th/0512172.

\bibitem[FV]{FV}  G. Felder, A. Varchenko, \textit{Multiplication formulas
for the elliptic gamma function}, math.QA/0212155.

\bibitem[KaZa]{KaZa}  M. Kaneko, D. Zagier, \textit{A generalized Jacobi
theta function and quasimodular forms}, in \textit{The moduli space of curves%
}, Proceedings of the Texel Island Meeting, April 1994, Birkh\"{a}user,
Basel 1995.

\bibitem[KO]{KO}  A. Kapustin, D. Orlov, \textit{Remarks on A-branes, mirror
symmetry, and the Fukaya category}, hep-th/0109098.

\bibitem[Kon]{Kon}  M. Kontsevich, \textit{Homological algebra of mirror
symmetry}, alg-geom 9411018.

\bibitem[KW]{KW}  A. Kapustin, E. Witten, \textit{Electric-magnetic duality
and the geometric Langlands program}, hep-th/0604151.

\bibitem[Pol 2003]{Pol 2003}  A. Polishchuk, \textit{Classification of
holomorphic vector bundles on noncommutative two-tori}, math.QA/0308136.

\bibitem[Pol 2004]{Pol 2004}  A. Polishchuk, \textit{Analogues of the
exponential map associated with complex structures on noncommutative two-tori%
}, math.QA/0404056v4.

\bibitem[Pol 2005]{Pol 2005}  A. Polishchuk, \textit{Quasicoherent sheaves
on complex noncommutative two-tori}, math.QA/0506571.

\bibitem[PS]{PS}  A. Polishchuk, A. Schwarz, \textit{Categories of
holomorphic vector bundles on noncommutative two-tori}, math.QA/0211062v2.

\bibitem[Rud]{Rud}  R. Rudd, \textit{The string partition function for QCD
on the torus}, hep-th/9407176.

\bibitem[Rui 1997]{Rui 1997}  S. N. M. Ruijsenaars, \textit{First order
analytic difference equations and integrable quantum systems}, J. Math.
Phys. 38 (1997), no.2, 1069-1146.

\bibitem[Rui 2001]{Rui}  S. N. M. Ruijsenaars, \textit{Special functions
defined by analytic difference equations}, in J. Bustoz, M. E. H. Ismail, S.
K. Suslov (eds.), \textit{Special functions 2000: Current perspective and
future directions}, NATO\ Sci. Ser. II Math. Phys Chem. 30, Kluwer,
Dordrecht 2001.

\bibitem[Soi]{Soi}  Y. Soibelman, \textit{Quantum tori, mirror symmetry and
deformation theory}, math.QA/0011162.

\bibitem[Sto]{Sto}  J. V. Stokman, \textit{Askey-Wilson functions and
quantum groups}, math.QA/0301330.

\bibitem[Wit]{Wit}  E. Witten, talk at workshop ``The Langlands program'',
CIRM\ Marseille, June 26th - July 2nd, 2006.
\end{thebibliography}
\end{document}